\documentclass[5p]{elsarticle}

\pdfoutput=1
\usepackage[]{hyperref}
\journal{Nuclear Instruments and Methods in Physics Research Section A: Accelerators, Spectrometers, Detectors and Associated Equipment}

\usepackage[separate-uncertainty = true]{siunitx}

\usepackage{epstopdf}

\usepackage{amsmath}

\begin{document}

\begin{frontmatter}

\title{Simulation Studies on Generation, Handling and Transport of laser-accelerated Carbon Ions}


\fntext[myfootnote]{Corresponding author, Tel.: +49 6159712748}

\author[TUDA]{J. Ding\fnref{myfootnote}}
\ead{jding@ikp.tu-darmstadt.de}
\author[GSI]{D. Schumacher}
\author[TUDA]{D. Jahn}
\author[GSI,JENA]{A.  Bla\ifmmode \check{z}\else \v{z}\fi{}evi\ifmmode \acute{c}\else \'{c}}
\author[TUDA,FAIR]{M. Roth}

\address[TUDA]{Institut f\"ur Kernphysik, Technische Universit\"at Darmstadt, Schlossgartenstrasse 9, D-64289 Darmstadt, Germany}
\address[GSI]{GSI Helmholtzzentrum f\"ur Schwerionenforschung, Planckstrasse 1, D-64291 Darmstadt, Germany}
\address[JENA]{Helmholtz Institut Jena, Helmholtzweg 4, D-07734 Jena, Germany}
\address[FAIR]{FAIR - Facility for Antiproton and Ion Research, Planckstrasse 1, D-64291 Darmstadt, Germany}
\begin{abstract}
To this day the interaction of high-intensity lasers with matter is considered to be a possible candidate for next generation particle accelerators. Within the LIGHT collaboration crucial work for the merging of a high-intensity laser driven ion source with conventional accelerator technology has been done in the past years. The simulation studies we report about are an important step in providing short and intense mid-Z heavy ion beams for future applications.
\end{abstract}

\begin{keyword}
Laser \sep Carbon \sep Ion \sep Acceleration
\end{keyword}

\end{frontmatter}


\section{Introduction}
Almost two decades went by since the discovery of compact acceleration of ions by means of ultra-intense laser irradiation of solid density targets \cite{TNSA_1,TNSA_2}, but still the mechanism did not fulfill its numerous promises. This new source for intense MeV-range ion bunches is still thought of as a driver for many applications. They range from igniting inertial confinement fusion \cite{Fusion_Laserprotons, Fusion_Laserprotons_2}, driving isochoric heating to achieve warm dense matter \cite{Isochoric_Heating_1, Isochoric_Heating_2, Isochoric_Heating_3} to the generation of isotopes with table-top devices \cite{table-top_isotopes} and particle therapy \cite{Particle_Therapy_1, RadioTherapyChallenge}. Due to the high particle numbers and initially short pulse durations of the ion bunches the monitoring of transient phenomena is an interesting field of work \cite{Proton_Radiographie_1}. Lately the generation of laser-driven neutron beams has been investigated by several groups \cite{Laser_neutrons_1} and neutron resonance spectroscopy was successfully demonstrated recently \cite{Laser_neutrons_2}.\\
Despite theoretical evidence in PIC simulations for potentially game-changing new acceleration mechanisms such as the Breakout-Afterburner (BoA) \cite{BoA_1} or the Radiation Pressure Acceleration \cite{RPA_1}, up to now the only reliable mechanism is Target Normal Sheath Acceleration (TNSA) \cite{TNSA_1,TNSA_2}. By irradiating micrometer thick targets with ultra-intense shortpulse lasers, which get focused down to more than \SI{e18}{\watt\per\square\centi\metre}, electrons from the front surface get pushed through the bulk material and form an electron sheath on the back side of the target. The resulting electrical field is of the order of \SI{e12}{\volt\per\metre}, ionizes the atoms on the rear surface and subsequently accelerates the ions up to several \SI{10}{\mega\electronvolt}. Based on the initially cold ion temperature, the absence of collisions during the acceleration and the process taking place on a picosecond time scale, the ion beams exhibit some exceptional characteristics. Total particle numbers exceed \SI{e12} and stem from a sub-mm source area with transverse emittances as low as \SI{0.01}{\milli\metre . \milli\radian}. While the energy spectrum of the ions follows an exponentially decaying distribution, there is a very high degree of order in the longitudinal structure of the beam, manifesting itself in the longitudinal emittance being below \SI{e-4}{\electronvolt\second} for specific energies \cite{Emittance_TNSA}. These characteristics make the TNSA a promising candidate for the next generation of accelerator technology \cite{Laserion_Applications_1}. While investigations of the underlying physics are well-advanced, the next step in coupling the laser-driven ion pulses to conventional accelerator structures is explored mainly in simulations. Among others the project Extreme Light Infrastructure (ELI) has dedicated a substantial working package, ELI-MED \cite{ELI-MED_1}, to the transport and handling of laser-driven ion pulses with ion optical systems.\\
For several years a collaborative effort has been undertaken by german universities and Helmholtz centres to examine \textit{L}aser \textit{I}on \textit{G}eneration, \textit{H}andling and \textit{T}ransport (\textit{LIGHT}) \cite{LIGHT_1}. The testbed is situated at GSI Helmholtz-centre, because it offers a high power high energy laser with PHELIX \cite{PHELIX_1} and accelerator expertise and infrastructure in the existing ion accelerator at GSI and the future FAIR facility.

\begin{figure*}[t]
\includegraphics[width=\textwidth]{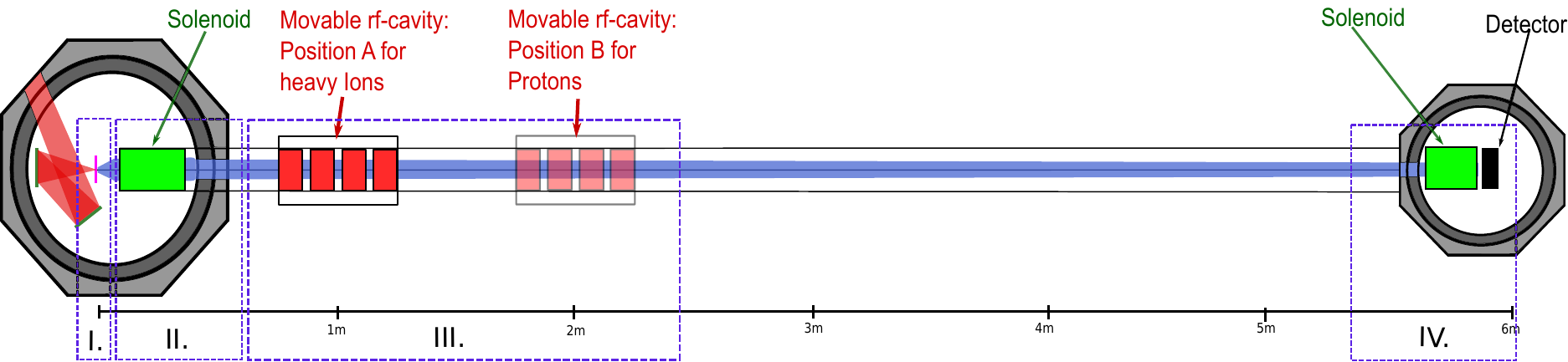}
\caption{The LIGHT beamline consists of a laser-driven ion source, based on the TNSA mechanism (I.), a collimation solenoid (II.), a cavity for energy compression or phase focussing (III.) and a second solenoid for final focussing in front of diagnostics (IV.). The cavity can be used at two different positions to enable optimal phase focussing.}
\label{LIGHT_Beamline}
\end{figure*}

\paragraph*{Status and achievements}
Work on the transport and focusing of laser accelerated protons was initiated in 2007 at Los Alamos National Laboratory, where permanent magnet quadrupole lenses were used to focus MeV protons from a TNSA source to sub-mm spots \cite{TNSA_Quad_1}. In 2008 the LIGHT program centred around transport and collimation of laser-accelerated proton beams was launched at GSI. After demonstrating the effectiveness of pulsed solenoids for the collimation of laser-accelerated proton beams \cite{Solenoid_Capture_1}, experimental campaigns were carried out in the proximity of the UNILAC at GSI. A sub-aperture beam of the PHELIX short pulse laser system can supply the experiment with up to \SI{25}{\joule} in \SI{650}{\femto\second} focused with an off-axis parabola to a spotsize of \SI{3.5}{\micro\metre} (FWHM). This leads to intensities exceeding \SI{e19}{\watt\per\square\centi\metre}. Protons generated via TNSA are captured by a pulsed solenoid and transported six metres to a diagnostics chamber (figure \ref{LIGHT_Beamline}). Two metres behind the target a radiofrequency cavity serves as buncher to decrease the energy spread or refocus the particles temporally. The efforts of the LIGHT collaboration to create short beams with high peak intensities culminated so far in the generation of proton beams with more than 10\textsuperscript{8} protons in \SI{0.5}{\nano\second} measured six metres behind the target \cite{LIGHT_3}.

\paragraph*{Objectives}
Many of the possible future applications of a laser-driven accelerator demand at least mid-Z ion beams. Therefore the next step in the evolution of laser-driven accelerators has to be the delivery of mid-Z ion beams with similar characteristics as it was managed with proton beams \cite{LIGHT_3}. The efficient acceleration of for example carbon ions has already been demonstrated successfully \cite{Laser_Carbonions_1}. Based on this work conducted also by members of our group, we propose to incorporate an improved target handling scheme in the LIGHT beamline. In order to replicate transport efficiency and temporal compression of the proton beams, simulation studies are necessary. Outcome and implications of these studies are presented in the following section.

\section{Simulation studies}
\label{sec:Simulation}
\paragraph*{Software} The simulations were carried out with the \textit{Trace-} \textit{Win} code \cite{TraceWin}. \textit{TraceWin} is primarily used for linear accelerator design and has features, which make the code suitable for simulating the transport of laser-accelerated ions. The code is a self-consistent 3D particle-in-cell code, which is capable of tracking up to 4$\cdot$10\textsuperscript{8} simulation particles. However it is not possible to create several beams, that act on each other, therefore a suitable space charge compensation (scc) factor is needed to include space charge for TNSA ion beams. \textit{TraceWin} can incorporate fieldmaps into the beamline and therefore allows for a more accurate simulation of ion optical elements and their aberrations.

\begin{figure}[!h]
  \includegraphics[scale = 1]{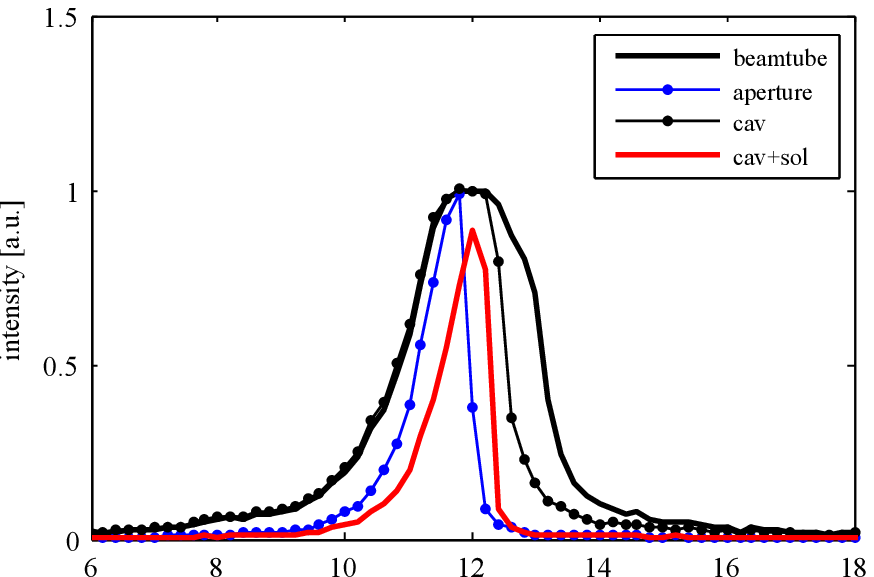}
  \includegraphics[scale = 1]{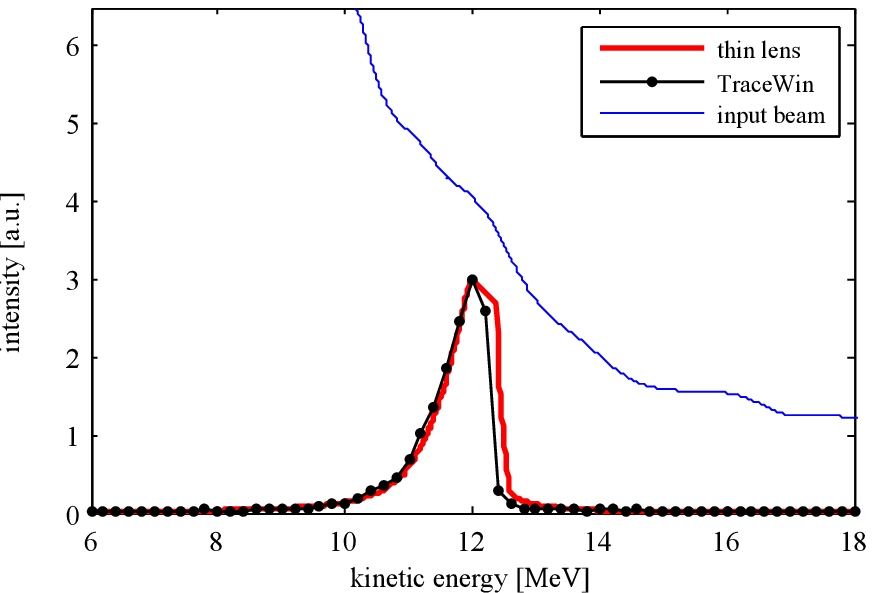}
  \caption{\textbf{Upper figure:} Simulations of laser-accelerated carbon ions with charge state 4+. For the same fieldmap of the solenoid four different beamlines were simulated. A \SI{100}{\milli\metre} diameter beam tube resulted in a spectrally broad transmission. By adding an aperture with \SI{20}{\milli\metre} diameter in between the solenoid and the monitor one reduces the spectral width of the transmission significantly. Taking the whole LIGHT-beamline setup with the apertures of a second solenoid and the cavity into consideration has almost the same effect on the energetic width of the resulting beam as the 20 mm aperture between solenoid one and detector.
  \textbf{Lower figure:} \textit{TraceWin} simulation of LIGHT-beamline in comparison with thin lens ray tracing approximation (both scaled by factor 3). On top of that the spectrum of the input beam is shown as well.}
  \label{fig:Transported_Carbon4}
\end{figure}

\paragraph*{Input}
In the presented simulations the input consisted of simulation particles, which were generated by choosing 6D phase space coordinates according to the measured source characteristics. The source characteristics were in parts taken from \cite{Laser_Carbonions_1,Laser_Carbonions_2}, from first experiments with heavy ions at LIGHT and from work done on laser accelerated proton beams. At an energy of \SI{12.1 \pm 0.1}{\mega\electronvolt} the input beam had an average divergence of \SI{197}{\milli\radian}, normalized transversal rms emittance of $\epsilon$\textsubscript{trans}=\SI{0.094}{\pi . \milli\meter . \milli\radian} and longitudinal rms emittance of $\epsilon$\textsubscript{long}=\SI{6.5e-8}{\pi . \eV . \second}. The whole input distribution had a bunch charge of \SI{6.41}{\nano\coulomb}, which equals 10\textsuperscript{10} C\textsuperscript{4+} ions with kinetic energies ranging from \SI{6}{\mega\electronvolt} to \SI{50}{\mega\electronvolt} (energy spectrum taken from heavy ion acceleration experiment). The beamline consisted of a solenoid \SI{0.04}{\metre} behind the target, a cavity (centre) at either \SI{2.275}{\metre} or \SI{1.275}{\metre} distance to the target and drift tubes up to a monitor at \SI{6.05}{\metre}. In figure \ref{LIGHT_Beamline} the LIGHT beamline is depicted with a second solenoid at the end of the beamline for final focusing. The solenoids were implemented as magnetic fieldmaps. All simulations were run with the \textit{Partran} distribution of \textit{TraceWin}, which tracks particles purely ballisticly. Space charge effects were included, assuming a radial symmetry. Initially the beam is quasi-neutral, because of co-moving electrons stemming from the TNSA mechanism. The magnetic fringe field of the first solenoid reflects the comoving electrons due to the magnetic mirror effect. Therefore simulations were carried out with a scc factor of 1 until after the solenoid, where the scc factor was set to 0.

\paragraph*{Outcome} In a first step carbon with charge state 4+ is simulated over \SI{6.05}{\metre} through a beam pipe with \SI{100}{\milli\metre} diameter. The fieldmap was set to emulate a solenoid from Hemholtz-centre Dresden-Rossendorf with a free aperture of \SI{40}{\milli\metre}, operated at a current of \SI{8.1}{\kilo\ampere} (maximum magnetic field and square of the magnetic field integrated on axis are \SI{6.72}{\tesla} and \SI{5.64}{\square\tesla \meter}). In a second simulation an aperture with \SI{20}{\milli\metre} diameter was added in the middle between the solenoid and the monitor. Comparing the result of the simulation with aperture with a simulation of the complete LIGHT-beamline setup (apertures of cavity and second solenoid equal \SI{35}{\milli\metre} and \SI{40}{\milli\metre} respectively) shows comparable energy widths with shifted peak energies (table \ref{tab:Energies}). The outcome of these transport simulations with C\textsuperscript{4+} is shown in the upper half of figure \ref{fig:Transported_Carbon4}. In the lower plot in figure \ref{fig:Transported_Carbon4} the input energy spectrum is depicted together with the LIGHT beamline output.
\begin{eqnarray}
	& \frac{1}{f} = \frac{q^2}{4 \gamma^2 m^2 v_z^2} \int B^2 \mathrm{d}z
\label{solenoid_lens}	
\end{eqnarray}
An estimation using a thin lens approximation of the sole- noid (equation \ref{solenoid_lens}) and ray tracing matches the output of \textit{TraceWin} well. Deviations of thin lens approximation and simulated beam only appear at the high energy cut-off of the transported beam and stem from space charge effects. For these energies, the particles are still diverging after the solenoid and the space charge pushes them even further out. The overall efficiency of the LIGHT beamline is \SI{19.7}{\percent} for the transport of laser-driven C\textsuperscript{4+} in the FWHM energy range from \SI{11.48}{\mega\electronvolt} to \SI{12.29}{\mega\electronvolt}. Going to lower solenoid currents would increase particle numbers because of the exponentially decreasing energy spectra of the input beam. Setting the peak energy to \SI{12.10}{\mega\electronvolt} is a trade-off between the energy of the ions and overall particle numbers especially with respect to phase focusing.

\begin{table}
\begin{center}
    \begin{tabular}{| l | c | c |}
    \hline
    & E\textsubscript{peak} [MeV] & E\textsubscript{width-FWHM} [MeV] \\ \hline
    beamtube - C\textsuperscript{4+} & 11.82  & 2.28 \\ \hline
    aperture - C\textsuperscript{4+} & 11.74 & 0.80\\  \hline
    cavity - C\textsuperscript{4+} & 11.82 & 1.73 \\  \hline
    LIGHT - C\textsuperscript{2+} & 2.20 & 0.28 \\  \hline
    LIGHT - C\textsuperscript{3+} & 6.43 & 0.79 \\  \hline
    LIGHT - C\textsuperscript{4+} & 12.10  & 0.81 \\  \hline
    LIGHT - C\textsuperscript{5+} & 19.12 & 1.14 \\  \hline
    \end{tabular}
\end{center}
    \caption{Peak energies and energy widths of carbon ions reaching the monitor at \SI{6.05}{\metre} for a solenoid pulsed with \SI{8.1}{\kilo\ampere}.}
    \label{tab:Energies}
\end{table}

\begin{figure}[h]
  \includegraphics[scale = 1]{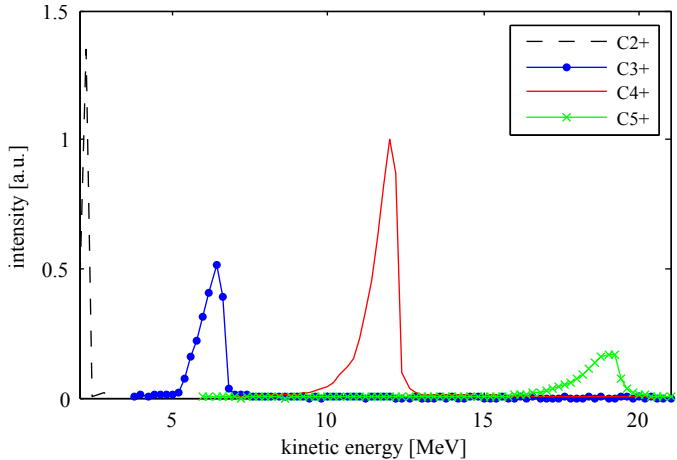}
  \caption{For different charge states, the focal lengths change according to equation \ref{solenoid_lens}. Therefore the lower the charge state, the lower the energy of the transported ions. The height of the peaks is normalized to the expected total particle number of C\textsuperscript{4+}.}
  \label{fig:Transported_Carbon}
\end{figure}

\paragraph*{Multiple charge states}
By using the TNSA mechanism to generate energetic ions one ends up with multiple charge states with very different energy spectra \cite{Laser_Carbonions_1,Laser_Carbonions_2} in one beam. Since the focal length of a solenoid depends on the charge to mass ratio (see equation \ref{solenoid_lens}) we expect peaks at different energies when looking at the transport of the complete beam. In figure \ref{fig:Transported_Carbon} the energy spectrum of a transport simulation of a carbon ion beam barring C\textsuperscript{1+} and C\textsuperscript{6+} is shown. Even if protons or C\textsuperscript{6+} ions were present, they would be strongly suppressed by the ion optical element. C\textsuperscript{1+} was transported at energies below \SI{1}{\mega\electronvolt}. The resulting peak energies and energy widths are written down in table \ref{tab:Energies}. The ratios of peak energies of different charge states follow a q\textsubscript{i}\textsuperscript{2}/q\textsubscript{j}\textsuperscript{2} law well. Abberations are to be attributed to different input distributions and to higher order effects.

\begin{figure}[htp]
  \includegraphics[scale = 1]{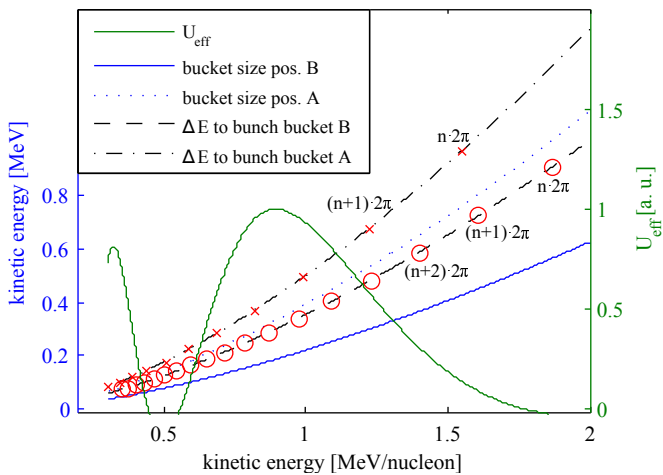}
  \caption{U\textsubscript{eff}, bucket size and necessary energy for successful bunching of a TNSA-generated C\textsuperscript{4+} beam are plotted against the kinetic energy of the particles injected at \SI{-90}{\degree} synchronous phase, both for position A and B of the cavity. The red $\circ$ and x are the velocity groups from equation \ref{eq:cavity_dv} (abscissa relevant only) arriving at the same phase at the cavity. For increasing kinetic energies the energy difference between them increases (more so for pos. A than for B).}
  \label{fig:Temporal_Focus}
\end{figure}

\paragraph*{Phase focusing}
Even though the transported ion bunches have a relatively small energy spread, they are still diverging longitudinally. For the purpose of either generating monoenergetic or longitudinally very short ion bunches a three gap spiral resonator was incorporated into the LIGHT beamline \cite{LIGHT_3, LIGHT_2} and demonstrated successfully for proton beams. To obtain similar results for at least one selected charge state one has to overcome two obstacles. Firstly the cavity consists of three gaps with drift tubes in between and the distances were designed to achieve the maximum effective gap voltage (U\textsubscript{eff}) for particles with \SI{8}{\mega\electronvolt}/u kinetic energies. With deviating kinetic energies U\textsubscript{eff} changes significantly (see figure \ref{fig:Temporal_Focus}). C\textsuperscript{4+} transport energy was set to around \SI{1}{\mega\electronvolt}/u to operate the cavity in a local maximum of U\textsubscript{eff}, which appear at $ v_{ion} = \frac{v_{design}}{2n+1}$. This also minimizes the sensitivity to small changes of the injection phase into the cavity. Secondly the time difference of the particles in the front and the back of the \SI{1}{\mega\electronvolt}/u C\textsuperscript{4+} beam exceeds the bucket size of the buncher cavity for the original position of the cavity (position B in figure \ref{LIGHT_Beamline}).

\begin{figure}[htp]
  \includegraphics[scale = 1]{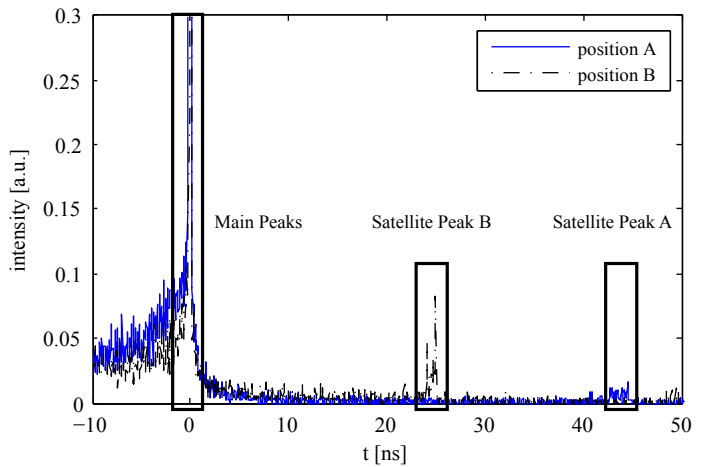}
  \caption{Normalized particle numbers plotted against time in case of bunching cavity at \SI{1.275}{\metre} (A) and \SI{2.275}{\metre} (B) distance from the target. The main peaks have an average energy of \SI{11.4}{\mega\electronvolt}.}
  \label{fig:Temporal_Focus_exp}
\end{figure}

 This can be improved by moving the cavity closer to the target to decrease the longitudinal extent of the beam at the position of the cavity. Experimental restrictions limit the minimum distance of the centre of the cavity to L=\SI{1.275}{\metre} (equals position A in figure \ref{LIGHT_Beamline}).
The buncher cavity was modeled as three thin gaps with transit time factor and drifts in between and F=\SI{108.4}{\mega\hertz} as radiofrequency. Figure \ref{fig:Temporal_Focus_exp} shows the outcome of simulations for both positons of the cavity. By optimizing the phase of injection into the cavity and the gap voltage we could demonstrate C\textsuperscript{4+} ion bunches with FWHM pulse durations of \SI{0.22}{\nano\second} for the cavity at position A, peak energy of \SI{11.38 \pm 0.06}{\mega\electronvolt}, and \SI{0.11}{\nano\second} at position B, peak energy of \SI{11.47 \pm 0.09}{\mega\electronvolt}. The temporally broader ion peak for the cavity at position A comes from the increased distance from cavity to monitor, and the focussing being less steep. In case A the satellite peak moved further away from the main peak, relating to equation \ref{eq:cavity_dv}, where n denotes the number of a velocity group in which particles can be bunched.
\begin{eqnarray}
	& v_n = \frac{FL}{n} \quad n= 1,2,...
	\label{eq:cavity_dv}
\end{eqnarray}
In a time bin of \SI{0.5}{\nano\second}, \SI{6.4}{\percent} of all input C\textsuperscript{4+} ions in the FWHM transport energy range were bunched together for position A (peak charge of \SI{15.0}{\pico\coulomb}) and \SI{4.1}{\percent} for position B (peak charge of \SI{9.6}{\pico\coulomb}). A real TNSA carbon ion source coupled with this beamline can therefore provide bunches with 10\textsuperscript{7} to 10\textsuperscript{8} particles in \SI{0.5}{\nano\second}. With a second solenoid the intense beam can be focussed down to below \SI{1}{\milli\metre} diameter (FWHM) \SI{0.1}{\meter} behind the second solenoid.

\section{Conclusion}
The simulation studies presented in this article on transport efficiency and generation of high intensity carbon ion bunches show, that with small changes in the existing LIGHT beamline setup C\textsuperscript{4+} ion bunches with up to 10\textsuperscript{8} particles in as short as \SI{0.5}{\nano\second} can be generated.

\section*{Acknowledments}
This work is supported by HIC4FAIR.

\bibliography{NIMA_PROCEEDINGS-D-17-00180_JDing}

\end{document}